\documentclass[12pt]{article}
\usepackage{amsfonts,latexsym,fullpage,amsmath,latexsym,amssymb,epsf}

\newtheorem{Theorem}{Theorem}

\newcommand{\onemat}[0]{{\mathbf 1}}

\newcommand{\cH}[0]{{\mathcal H}}

\newcommand{\R}[0]{{\mathbb{R}}}

\newcommand{\qed}[0]{\hfill $\Box$}

\title{\Large \textbf{Bounds on the number of time steps 
for \\ simulating arbitrary interaction graphs}} \author{ Dominik
Janzing\thanks{e-mail: {\protect\tt
\{janzing,wocjan,eiss\_office\}@ira.uka.de}}, Pawe{\l} Wocjan, and Thomas
Beth \\ \small Institut f{\"u}r Algorithmen und Kognitive Systeme,
Universit{\"a}t Karlsruhe,\\[-1ex] \small Am Fasanengarten 5,
D-76\,131 Karlsruhe, Germany}

\date{September 26, 2002}

\begin{document}

\maketitle

\abstract{In previous papers we have considered
mutual simulation of $n$-partite pair-interaction
Hamiltonians. We have focussed on the {\it running time
overhead} of general simulations, while considering the
required {\it number of time steps} only for special cases
(decoupling and time-reversal). 
These two complexity measures differ significantly. Here we derive 
lower bounds on the number of time steps for general simulations.
In particular, the simulation of interaction graphs with irrational spectrum
requires at least $n$ steps. 
We discuss as examples graphs that correspond to graph codes
and nearest neighbor interactions in $1$- and $2$-dimensional lattices.
In the latter case the lower bounds are almost tight.
}

\section{Introduction}
Simulating Hamiltonian evolutions of arbitrary quantum systems on a
quantum computer is an idea that goes back already to Feynman
\cite{feynman}. This
would be an attractive application of future quantum computers
since there are no known efficient classical algorithms
for simulating generic dynamics of many particle systems. 
Here a quantum computer can be any quantum system provided that its
time evolution can be controlled in a universal way.
In particular, the problem of controllability includes the question which 
Hamiltonian evolutions can be simulated efficiently by the
considered Hamiltonian system [2--13]
Assume
the natural Hamiltonian $H$ to act on an $n$-fold tensor product
Hilbert space
\[
\cH_n := \cH\otimes \cH \otimes \dots \otimes \cH
\]
where each $\cH$ denotes the Hilbert space of a qudit, i.e. a
$d$-dimensional  system. The Lie algebra $su(d)$ of traceless Hermitian
operators on $\cH$ is a $m:=(d^2-1)$-dimensional real vector
space. Let $B:=\{\sigma_\alpha\mid \alpha=1,\ldots,m\}$ be an
orthogonal basis of $su(d)$ with respect to the trace inner product
$\langle A|B \rangle:=tr(A^\dagger B)/d$ for $A,B\in su(d)$.

Then the most general qudit-qudit interaction on $n$ coupled qudits is
given by
\[
H_J:=\sum_{k<l} \sum_{\alpha\beta} 
J_{kl;\alpha\beta} \sigma_\alpha^{(k)} \sigma_\beta^{(l)}\,,
\]
where $J$ is chosen to be a real symmetric $mn \times mn$-matrix with
zeros for $k=l$. Note that the symmetry of the coupling matrix $J$
does not imply any physical symmetry of the interaction. It is a
consequence of our redundant representation that turns out to be very
useful. The coupling matrix $J$ consists of $m\times m$-blocks. The
$m\times m$-matrix $J_{kl}$ given by the block at position $(k,l)$
describes the coupling between the qudits $k$ and $l$. We have
$J_{lk}=J_{kl}^T$, i.\,e. the matrix describing the coupling between
the qudits $l$ and $k$ is just the transpose of the matrix describing
the coupling between $k$ and $l$. The blocks on the
diagonal are zero matrices.

In the setting discussed here and in most other articles on simulation
of Hamiltonians the only possibilities of external control are given by
local unitaries on each qudit. We assume that one is able to implement
them independently. Formally, all control operations are elements of
the group ${\mathcal C}:=SU(d)\otimes SU(d)\otimes\cdots\otimes
SU(d)$. A common approximation is to assume that all operations in
${\mathcal C}$ can be implemented arbitrarily fast (``fast control
limit''). The simulation of Hamiltonians is based on the following
``average Hamiltonian'' approach 
 that has successfully been used 
for describing Nuclear Magnetic Resonance experiments since many years 
(e.g. \cite{slichter,rhim}).

 Let $t_1,t_2,\dots,t_n$ be positive
numbers and $u_1,u_2,\dots,u_N \in {\mathcal C}$ be control
operations. Then the algorithm
\begin{quote} 
{\em perform $u_1$, wait $t_1$, perform $u_1^\dagger$,
perform $u_2$, wait $t_2$, perform $u_2^\dagger$, 
\ldots,
perform $u_N$, wait $t_N$, perform $u_N^\dagger$}
\end{quote}
implements approximatively the unitary evolution
\[
\prod_{j=1}^N \exp (u^\dagger_j H u_j t_j  )\,.
\]
If the times $t_j$ are small compared to the time scale of the natural
evolution according to the natural Hamiltonian 
$H$ this is approximatively the evolution
according to the {\it average Hamiltonian}
\begin{equation}\label{average}
\sum_j t_j u_j H u_j^\dagger / \tau \,,
\end{equation}
where $\tau:=\sum_j t_j$ is the slow down factor of teh evolution, 
i.e., the 
{\it time overhead} 
of the simulation.
For investigating the time overhead and the number of time steps of a
simulation it turns out to be useful to work with the coupling matrices
instead of considering the Hamiltonians themselves.
To express the effect of the
control operations on the coupling matrix $J$ note that any unitary
operation $u\in SU(d)$ corresponds to a rotation on the
$m$-dimensional sphere via the relation
\[
u^\dagger \big( \sum_\alpha c_\alpha \sigma_\alpha \big) u = 
\sum_\alpha \tilde{c}_\alpha \sigma_\alpha\,,
\]
where the vector
$\tilde{c}=(\tilde{c}_1,\tilde{c}_2,\ldots,\tilde{c}_m)$ is obtained
by applying a rotation $U \in SO(m)$ on the vector
$c=(c_1,c_2,\ldots,c_m)$. It is straightforward to verify that
conjugation of $H_J$ by $v:=u^{(1)}\otimes u^{(2)}\otimes\cdots\otimes
u^{(n)}$ corresponds to conjugation of $J$ by a block diagonal matrix
of the form
\[
V:=U^{(1)}\oplus U^{(2)}\oplus\cdots\oplus U^{(n)}\in 
\bigoplus_{k=1}^n SO(m)\,.
\]
The condition for correct simulation is hence given by
\begin{equation}\label{condition}
\tilde{J}= \sum_j t_j V_j J V_j^T \,,
\end{equation}
where the orthogonal matrix $V_j$ corresponds to the unitary $v_j$ for
$j=1,\ldots,N$.

The question of optimal simulation has been completely solved so far
only for the case of two-qubit Hamiltonians
\cite{bett,vidalcirac}. Optimal simulation
protocols are constructed  in \cite{vidalcirac}. The number of time steps is
between $1$ and $3$.

For $n$ qudits we have shown in
\cite{graph,arrow,efficient} that the eigenvalues of the
coupling matrix $J$ and $\tilde{J}$ provide {\it lower} bounds on the
simulation time overhead. Graph theoretical notions can also provide
{\it upper} bounds. In certain cases the bounds are known to be tight
\cite{graph,arrow}. 

Here we do not focus on the time overhead but on the number of time
steps in the $n$ partite case. Upper bounds for decoupling (switching
off the Hamiltonian) and time-reversal (simulating of $-H$ by $H$) are
directly given by the parameters of known schemes
\cite{leung,stoll,rev,efficient}. Lower bounds 
for time-reversal have been derived 
in \cite{arrow}. These results shown that time overhead and the number
of time steps are not connected in any obvious way.
Here we address the general problem of mutual simulation of
pair-interactions on $n$ qudits. 

In Section~2 we derive lower bounds on the number of time steps
for the general problem of mutual simulation of pair-interactions.
The type of  
coupling is assumed to be the same between all nodes, but the strengths
and the signs may vary.
Like the bounds on the time overhead \cite{graph,bett,efficient}, 
the bound derived here make us of the spectrum
of the corresponding coupling matrices.  
In the special case that one wants to cancel some interactions while keeping 
others (``the simulation of a certain interaction graph'')
the bounds refer directly to the spectra of the corresponding adjacency 
matrices.
Surprisingly, the greatest lower bound can be given if the smallest eigenvalue
of the adjacency matrix is irrational.
In Section 3 the bounds shall be applied to three different cases
where long range interactions are present between all spins.
The first two cases are the simulation of nearest neighborhood coupling 
in a square lattice  and a cyclic chain. The third case
is a certain interaction graph that has been proposed 
to prepare the states of a graph code \cite{GraphCodes1,GraphCodes2,GrRoe}.

\section{Lower bounds on the number of time steps}
We restrict
our attention to interactions with an additional symmetry, namely 
Hamiltonians of the following form
\begin{equation}\label{weightedGraph}
H:=\sum_{k<l} w_{kl} \sum_{\alpha\beta} c_{\alpha\beta}
\sigma_\alpha^{(k)}\sigma_\beta^{(l)} \,.
\end{equation}
The matrix $W:=(w_{kl})$ is a real symmetric $n\times n$-matrix with
zeros on the diagonal. It describes the coupling strengths and the
signs of the interactions between all qudits. The matrix
$C=(c_{\alpha\beta})$ is a real symmetric $m\times m$-matrix
characterizing the type of the coupling. This means that all qudits
interact with each other via the same interaction and that only the
coupling strengths and the signs vary. It is important that in this
special case the coupling matrix $J$ can be expressed as a tensor
product of $W$ and $C$, i.\,e., $J = W \otimes C$.

To derive a general lower bound for simulating arbitrary 
interactions $\tilde{J}$ by a tensor product interaction
$J=W\otimes C$ it is useful to observe that the decisive condition
in eq.~(\ref{condition}) is invariant with respect to the following 
{\it rescaling} of interactions: Multiply each $m\times m$-block $k,l$
of $J$ and $\tilde{J}$ with the same factor $r_{kl}$.
In the case $W\otimes C$ we can therefore  assume w.l.o.g. that
$W$ is a matrix with only $1$ as non-diagonal entries
as long as we restrict our attention to Hamiltonians with complete
interaction graphs, i.e., $w_{kl}\neq 0$ for $k\neq l$.
Formally, rescaling is denoted as follows. Let $A/B$ be the entry-wise
quotient of the matrices $A$ and $B$ provided that $A$ has only
zero entries at those positions where $A$ has also a zero.
Then we consider the simulation
of $\tilde{J}/ (W\otimes I)$ by $K\otimes C$
where $K$ is the matrix with only one as non-diagonal entries
and $0$ in the diagonal and $I$ is the matrix with only $1$ as entries.

In order to emphasize that the assumption $w_{kl}=1$  is not 
an assumption on the real physical coupling strength we
formulate the following theorem for general $W$ and use rescaling
only in the proof.

\begin{Theorem}[Lower bound]\label{Haupt}
Let $J:=W\otimes C$ 
be the coupling matrix of the system Hamiltonian, $\tilde{J}$ an  arbitrary
coupling matrix of the interaction that is simulated, and
$\mu$ the time overhead. Denote the minimal and maximal eigenvalues of $C$ by
$\lambda_{\min}$ and $\lambda_{\max}$, respectively
and its rank by $r(C)$. Let $I$ be the
$m\times m$-matrix whose all entries are $1$.
Let $s$ be the number of eigenvalues of $\tilde{J}/(W \otimes I)$ that
are not contained in the interval
\[
{\mathcal I}:=[\,-\mu \lambda_{\max}, -\mu \lambda_{\min}\,]\,.
\]
Then the number of time steps required to simulate $H_{\tilde{J}}$ by $H_J$
is at least $s/r(C)$.
\end{Theorem}

\vspace{0.3cm}
\noindent
{\it Proof:} The condition for a scheme
$t_1,V_1,t_2,V_2,\dots,t_N,V_N$ to be a simulation of $\tilde{J}$
reads
\begin{equation}\label{cond}
\sum_{j=1}^N t_j V_j (W\otimes C)  V_j^T=\tilde{J}\,.
\end{equation}
Since the matrices $V_j$ are block-diagonal we can rescale each
$m\times m$-block such that we obtain
\[
\sum_{j=1}^N t_j V_j (K \otimes C) V_j^T =\tilde{J}/ (W\otimes I)\,.
\]
In the following we denote the rescaled coupling matrix
$\tilde{J}/(W\otimes I)$ by $J'$.

Set $R:=\sum_{j=1}^N t_j V_j ({\mathbf 1}\otimes C) V_j^T$. By adding
$R$ on both sides we obtain
\begin{equation}\label{specialform}
\sum_{j=1}^N t_j V_j \big((K+{\mathbf 1})\otimes C\big) V_j^T = J' +
R\,.
\end{equation}
The rank of the matrix $(K+{\mathbf 1})$ is $1$ since all its entries
are $1$. Consequently, the rank of the left hand side of
eq.~(\ref{specialform}) is at most $N\, r(C)$. 

Set $a:=\mu \lambda_{\min}$ and $b:=\mu \lambda_{\max}$.
Note that the eigenvalues of $R$ are contained in the
interval $[a,b]$. 
The rank of the
right hand side is at least the number of eigenvalues of $J'$ outside
the interval ${\mathcal I}=[-b,-a]$. This is seen as follows.

Let $P$ and $Q$ be the projections onto the sums of all
eigenspaces of $J'$ with eigenvalues smaller than $-b$ and greater than
$-a$, respectively. Clearly, $s$ is equal to the dimension of the image of
$P\oplus Q$. 
Denote by $\hat{J}$ and $\hat{R}$ the $s\times
s$-submatrices of $J'$ and $R$ defined by $(P\oplus Q) J' (P\oplus Q)$
and $(P\oplus Q) R (P\oplus Q)$, respectively.

Due to the choice of $P$ and $Q$ the spectrum of $\hat{J}$ is
contained in the interval $(-\infty,-b)\cup (-a,\infty)$. The spectrum
of $\tilde{R}$ is contained in the interval $[a,b]$ since the minimal
(maximal) eigenvalues of a matrix cannot decrease (increase) when
projecting the matrix.

We prove the theorem by showing that $\hat{J}+\hat{R}$ has full
rank. Set $f:=\frac{1}{2}(a+b)$. From the triangle inequality
it follows that for every unit vector $|\Psi\rangle \in \R^s$ one has
\begin{eqnarray*}
\|(\hat{J}+\hat{R})|\Psi\rangle\| & = &
\|(\hat{J}+f\onemat+\hat{R}-f\onemat)|\Psi\rangle\|
\\
& \ge &
\|(\hat{J}+f\onemat)|\Psi\rangle\|-
\|(\hat{R}-f\onemat) |\Psi\rangle\|\,.
\end{eqnarray*}
The eigenvalues of the shifted operators $\hat{J}+f\onemat$
and $\hat{R}-f\onemat$ are contained in the intervals
$\big(-\infty,-\frac{1}{2}(b-a)\big)\cup\big(\frac{1}{2}(b-a),\infty\big)$
and $[-\frac{1}{2}(b-a),\frac{1}{2}(b-a)]$, respectively. This implies
that
\[
\|(\hat{J}+f\onemat)|\Psi\rangle\|-
\|(\hat{R}-f\onemat) |\Psi\rangle\| > 0
\]
because the norms can be bounded by the absolute values of the
eigenvalues:
$\|(\hat{J}+f\onemat)|\Psi\rangle\|>|\frac{1}{2}(b-a)|$ and
$\|(\hat{R}-f\onemat)|\Psi\rangle \|\le|\frac{1}{2}(b-a)|$.
This completes the proof. \qed

\vspace{0.3cm}

The following upper bound can be easily derived from 
Caratheodory's theorem \cite{Rock}.

\begin{Theorem}[Upper bound]
Every simulation that is possible can be achieved within
\[
\frac{n(n-1)}{2} m^2 +1
\]
time steps.
\end{Theorem}

\noindent
{\it Proof}\,\,\,
If $\tilde{J}$ can be simulated by $J$ with time overhead $\tau$ then
$\tilde{J}/\tau$ is in  the convex span of the matrices 
$V_j J V^T_j$ with notation as above. The dimension of this convex set
is at most $m^2 n(n-1)/2$ since the diagonal blocks are empty
and each matrix $V_j JV_j^T$ is symmetric.
Caratheodory's theorem states that each point in an $M$ dimensional
convex set can be written as a convex sum of at most $M+1$ extreme
points. $\Box$
\vspace{0.4cm}

There are interesting cases where the bound of Theorem \ref{Haupt} 
can be tightened. Assume $\tilde{J}=\tilde{W}\otimes C$ with the 
same matrix $C$ as the interaction that is used for the simulation.
In other words, only the strengths and the signs of some interactions should 
be changed. By the same rescaling trick as above we can consider 
the problem to simulate $(\tilde{W}/W)\otimes C$ by $K\otimes C$ where
$K$ has only $1$ as non-diagonal entries.

\begin{Theorem}[Lower bound]\label{Haupt2}
Let  $W\otimes C$ be the coupling matrix of the natural Hamiltonian
and $\tilde{W}\otimes C$ the coupling that we want to simulate. 
Assume all non-diagonal entries of $W$ to be non-zero.

\begin{enumerate}
\item
Let $C$ be a positive semidefinite  matrix. Then the number of time steps
is at least the number of positive eigenvalues of $\tilde{W}/W$.
 
\item 
Let $C=diag(1,1,\ldots,1)$ be the $m\times m$-identity matrix. Then
the number of time steps is at least $n-k$, where $k$ is the
multiplicity of the smallest eigenvalue $\mu_{\min}$ of $\tilde{W}/W$.

\item 
Let the natural coupling be $C:=diag (0,0,1)$, i.e., we have 
$\sigma_z\otimes \sigma_z$ interactions between all spin-1/2-particles.
  Let the set of local control operations be restricted
to $i\sigma_x$--transformations.
Then one requires at least $n-k$ time steps with $k$ as in Case~2.
If $\mu_{\min}$ is irrational then at least $n$ steps are  necessary.
In any case, 
$n(n-1)/2+1$ time steps are always sufficient. 

\end{enumerate}

\end{Theorem}

Note that Case (3) is of special interest since it deals with simulation 
procedures that do not rely on any first order approximation. 
In this case all  summands in eq.~(\ref{average}) commute and the 
unitary transformation implemented by the simulation scheme
coincides exactly with the exponent of the average Hamiltonian.

\vspace{0.3cm}

\noindent
{\it Proof} \, (of Theorem   \ref{Haupt2}) \,\,

\noindent
{\bf Case 1}
This statement is a corollary of Theorem \ref{Haupt}
since $\tilde{J}/(W\otimes I)=(\tilde{W}/W)\otimes C$.
The number of positive eigenvalues of this tensor product matrix
is $r(C)$ times the number of positive eigenvalues   
of $\tilde{W}/W$ if $r(C)$ is the rank of $C$.

\vspace{0.2cm}

\noindent
{\bf Case 2}
Consider the rescaled problem to simulate
$A\otimes C$ by $K\otimes C$.
In this case the right hand side of eq. (\ref{specialform}) 
reduces to 
\begin{equation}\label{Ausdr}
A \otimes C + \tau {\mathbf 1}\,,
\end{equation}
where $\tau:=\sum_j t_j$ is the time overhead and $A:=\tilde{W}/W$.
In \cite{efficient} we have shown that $\tau$ is at least
$-\mu_{\min}$.
 The reason is that 
the spectrum of $\tau K$ has to {\it majorize} the spectrum of
$A$ since $A$ is a convex sum of conjugates of $\tau K$. 
 Therefore the rank of the matrix in expression
(\ref{Ausdr}) is at least $m(n -k)$.  Since the left hand side of
eq.~(\ref{specialform}) has at most the rank $m\,N$ the number of time
steps is at least $n-k$.

\vspace{0.2cm}
\noindent
{\bf Case 3} In this case one can use a more convenient formalism.  We
drop the matrix $C$ and characterize the interactions by $W$ and
$\tilde{W}$. If a $zz$-interaction is conjugated by
$i\sigma_x$-transformations only two possibilities  occur. The
interaction term between spin $k$ and spin $l$ acquires a minus sign
if exactly one spin of both is subjected to a conjugation. The
interaction is unchanged if either no spin 
or both spins are conjugated. Instead
of representing the time step $j$ by the $3n\times 3n$ block diagonal
matrix $V_j$ we can represent it by an $n\times n$ diagonal matrix
$X_j$. The diagonal entries are $\pm1 $ and indicate which spins are
subjected to conjugation. Then eq.~(\ref{cond}) reduces to
\[
\sum_{j=1}^N t_j X_j K X_j= A\,.
\]
We add the identity matrix on both sides.
Due to $X_j {\mathbf 1} X_j ={\mathbf 1}$ we obtain
\[
\sum_j t_j X_j (K+{\mathbf 1}) X_j = A+ \sum_j t_j {\mathbf 1}\,.
\]
The rank of the left hand side is at most the number $N$ of time
steps.  To estimate the rank of the right hand side note that the time
overhead $\tau:=\sum_j t_j$ is at least $-\mu_{\min}$. Hence the dimension of the kernel
of $A+\tau {\mathbf 1}$ can at most be the multiplicity of the eigenvalue
$\mu_{\min}$.  This shows that the number of time steps is at least
$n-k$.

Assume $\mu_{\min}$ to be irrational. Then the time overhead 
$\tau$ is necessarily greater than
$-\mu_{\min}$. This can be seen as follows.  The optimization with
respect to the time overhead reduces to the following convex problem.
Consider the matrix $X_j K X_j$ for an arbitrary time step $j$.
Its non-diagonal entries
are $\pm 1$ and indicate which interactions acquire a minus
sign in $j$th step. 
In graph-theoretical language,  the set of matrices that can occur as
$X_j KX_j$ are exactly the
Seidel matrices of complete bipartite graphs
(see \cite{graph}). 
Then the optimal $\tau$ is the minimal positive
number such that $A/\tau$ is in the convex span of the set of Seidel
matrices of complete bipartite graphs.  Geometrically, the convex span
is a polytope having the Seidel matrices as its extreme points. It is embedded
in the $n(n-1)/2$ dimensional vector space of real symmetric matrices
with zeros on the diagonal. Let $O$ be the origin. Consider the semi-line
$\nu A$ for $\nu\ge 0$. Then the optimal simulation is the unique
intersection point $P$ of the semi-line with the boundary of the
polytope. The quotient of the distance between $0$ and $A$ and between
$0$ and $P$ is the optimal time overhead. This quotient can never be
irrational. The reason is that $P$ has rational entries since it is
the solution of a linear equation over the field $\mathbb{Q}$
of rational numbers. Hence
$\tau$ is greater than $-\mu_{\min}$ and $A+\tau\mathbf{1}$ has
necessarily full rank. This proves that we need at least $n$ time
steps.

To see that $n(n-1)/2 +1$ time steps are always sufficient we can
argue as in the proof of Theorem \ref{Haupt} with Caratheodory's
theorem. The dimension of the convex span of the matrices $X_j K X_j$
is at most $n(n-1)/2$. \hfill $\Box$

\vspace{0.4cm}

In the following section we will consider the task to cancel 
some interactions and keep others. Then $A:=\tilde{W}/W$ has only $1$ 
and $0$ as entries. In graph-theoretical language, it is
the adjacency matrix of the desired interaction graph.

It is surprising that it is relevant for our lower bound
whether the smallest eigenvalue of the adjacency matrix is irrational
(see Case 3.).
 It is not clear whether this is 
only a feature of our proof or whether there is a true connection
to the irrationality of graph spectra. 

In the following section we will use graph spectra for deriving
lower bounds based on  the above theorems. We show that some of them are 
almost tight by
sketching
simulation schemes based on well-known results on selective decoupling.

\section{Applications}

In this section we consider the simulation of special interaction graphs
that appeared in the literature in various  applications.
Some
interesting models in quantum information theory refer to quite
idealized types of Hamiltonians like nearest neighbor
interactions. If the natural Hamiltonian contains long range
interactions between all nodes one may try to simulate the idealized
interaction. Then the problem is to cancel the unwanted terms without
destroying the desired interactions.  The examples below show that it
may cause a large number of time steps to cancel unwanted long-range
interactions no matter how fast they are decreasing with the
distance. As long as they are not neglected, the control sequences
that cancel them may be rather long.  Here we restrict our attention
to $\sigma_z\otimes\sigma_z$ interactions between $n$ qubits.

\subsection{Cyclic and open spin chains}

\begin{figure}\label{chains}
\centerline{
\epsfbox[-10 -10 160 20]{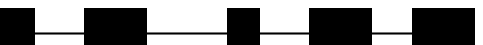} 
\epsfbox[0 0 140 70]{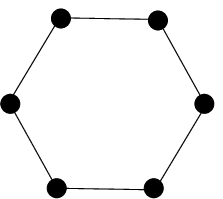}
}
\caption{{\small Open and cyclic spin chains}} 
\end{figure}

We examine first a quantum system of $n$ spins equally spaced on a
one-dimensional lattice. The interaction between the spins may decrease
with the distance between the spins. We consider the problem to
simulate a chain with only nearest neighbor interactions, i.e. we have
to cancel the interactions between all non-adjacent pairs.
The desired interaction graphs can be seen in Fig.~\ref{chains}.

The adjacency matrix  of the cyclic spin chain 
is $S+S^T$, where $S$ denotes the cyclic shift in $n$ dimensions.
The eigenvalues of $S$ and $S^T$ are the $n$th roots of unity.
The 
eigenvalues of $S+S^T$ are given by
\[
2 \Re \Big(\exp(i2\pi l/n)\Big) \,\,\,\hbox{ with } \,\,\,0\leq l < n \,.
\]

For odd $n$ the lower bound on the number of time steps is $n$ since 
the least eigenvalue is irrational for all $n>3$. 
If $n$ is even it is rational and has 
multiplicity $1$. Hence the bound is $n-1$ in this case.
There are numbers $n$  where this bound is almost tight. This is shown by 
the
following example.
Let 
$n$ be an even number with the property that a Hadamard matrix
of dimension $n/2$ exists. This is for instance the case
for each power of $2$  (see \cite{sloane}).
We construct a simulation scheme
that consists of $2$ subroutines.

The first subroutine 
simulates the interaction between the pairs
$\{1,2\},\{3,4\},\dots,$ $\{n-1,n\}$
and the second simulates $\{n,1,\},\{2,3\},\dots,\{n-2,n-1\}$.
Note that all the pairs in the same subroutine are disjoint.
The problem to simulate the interactions  between
disjoint pairs is a special instance of well-known ``cluster decoupling''
\cite{efficient}
where the interaction between independent cliques
are cancelled and the interactions within the same clique remains.
It can be achieved using Hadamard matrices having
at least the number of cliques as dimension (compare \cite{stoll,leung}).
The entries $\pm 1$ in column $j$ determine which spins are conjugated
in the step $j$. The dimension is the number of time steps of the 
decoupling subroutine.

Using this method, we need $n/2$ steps in each subroutine. Therefore we
have given a simulation with $n$ steps.
whereas our lower bound is $n-1$. 
In general the number of steps for simulating the circle
grows only linearly in $n$. This shows that the lower bound is quite good
even for general $n$.

The adjacency matrix corresponding to the open spin chain
has the eigenvalues \cite{CDS}
\begin{equation}
2\cos\Big(\frac{\pi}{n+1}\, i\Big)\,,\quad i=1,\ldots, n\,.
\end{equation}
The smallest eigenvalue is irrational for all $n>2$. 
By Theorem \ref{Haupt2} (Case 3) 
we conclude that
the number of time steps is at least $n$.

\subsection{Square lattice}
We consider a quantum system of $n=l^2$ spins located on a
two-dimensional square lattice. For simplicity 
assume $l$ to be even. We want to simulate a 
lattice with only nearest neighbor
interactions.

\begin{figure}\label{Gitt}
\centerline{
\epsfbox[0 0 355 66]{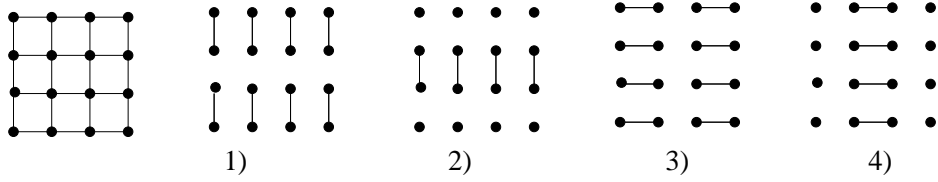} 
}
\caption{{\small Simulation of the square lattice interaction with $4$ subroutines}}
\end{figure}

The desired interaction graph is shown on the left of Fig.~\ref{Gitt}.
This kind of interaction can for instance be used for
preparing the initial state in the `One-Way Quantum Computer'
proposed in \cite{RB00}.
The eigenvalues of the corresponding adjacency matrix $A$ are 
known  in graph theory \cite{CDS}: 
\begin{equation}
2\cos\Big(\frac{\pi}{l+1}\, i\Big) + 2\cos\Big(\frac{\pi}{l+1}\,
j\Big) \,,\quad i,j =1,\ldots, l\,.
\end{equation}
We first consider the time overhead. An upper bound is given by
$4$ since this is  the
chromatic index of the graph \cite{graph}. 
It is easy to see that the minimal eigenvalue of $A$ is
given by
\begin{equation}
\lambda_{\min} = 
2\cos\Big(\frac{\pi}{l+1}\, l\Big) + 2\cos\Big(\frac{\pi}{l+1}\, l\Big)\,.
\end{equation}
By Theorem \ref{Haupt2} (Case 3) the lower bound on 
the number of time steps is $n$ 
since the smallest eigenvalue is irrational.
Note that this example shows that the complexity measures
{\it time overhead} and {\it number of time steps} may differ
significantly. 

An upper bound on the number of time steps can be obtained as follows.
The graph  has $2(l-1)l$ edges. We can partition
the edges into $4$ sets of edges such that each set contains
only disjoint interacting pairs.
These $4$ partitions  are shown in Fig.~\ref{Gitt}.
The simulation consists of $4$ subroutines simulating one the interactions
in one of the $4$ classes.
For each subroutine we choose Hadamard matrices with a dimension 
that is at least the number of cliques.
The numbers of cliques are $l^2/2$ or $l^2/2+l$ in each subroutine. 
Since there exist Hadamard matrices for every power of $2$ 
the square lattice graph can always be simulated in
$O(l^2)=O(n)$ time steps.

\subsection{Graph codes}
\begin{figure}\label{GR}
\centerline{
\epsfbox[0 0 120 90]{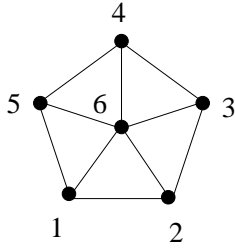} 
}
\caption{{\small Interaction required for preparing
states of a graph code of length 5.}}
\end{figure}

The computational power of different $n$ spin interactions is not well
understood yet. It would be interesting to know which $n$ qubit
transformations can easily be implemented when a certain Hamiltonian
is given.  However, one of the few examples where the power of
specific Hamiltonians is {\it directly} used (without using them to
implement 2-qubit gates) is the preparation of states of graph codes
proposed in \cite{GraphCodes1,GraphCodes2}. 
Here the codes states are obtained by the free
time evolution according to a Hamiltonian with
$\sigma_z\otimes\sigma_z$ interactions. The graph representing 
a code is the interaction graph that can be used
for preparing the states.
We assume the natural interaction to be
equal $zz$ interaction between all $6$ spins.
and would like to simulate the interaction graph in Fig~\ref{GR}.

The eigenvalues of the `wheel' in Fig.~\ref{GR}  can easily be computed by
any computer algebra system.  They are 
\[
1+\sqrt{6}\,,\,\,\, 
\frac{1}{2}\sqrt{5}-\frac{1}{2}\,,\,\,\, 
\frac{1}{2}\sqrt{5}-\frac{1}{2}\,,\,\,\,
1-\sqrt{6}\,,\,\,\,
-\frac{1}{2}\sqrt{5}-\frac{1}{2}\,,\,\,\,
-\frac{1}{2}\sqrt{5}-\frac{1}{2}\,.
\]

The minimal eigenvalue is $-1/2 -\sqrt{5}/2$. 
By Theorem \ref{Haupt2} (Case 3) the minimal number of time steps is $6$.
An implementation with $12$ time steps  is given as follows.
The scheme consists of $3$ subroutines each consisting of $4$ 
time steps. As above each 
subroutine simulates the interaction between 
disjoint cliques and cancels the interaction between different cliques. 
Subroutine $1$ has the cliques $\{1,2,6\},\{4,5\},\{3\}$.  
The clique partitions in subroutine $2$ and $3$  are
$\{3,4,6\},\{1,5\},\{2\}$ and $\{1\},\{4\},\{2,3\},\{5,6\}$, respectively.
In each subroutine
decoupling the different cliques can be achieved by
Hadamard matrices of dimension $4$ since no subroutine
has more than $4$ cliques. Hence we have $4$ time steps in each subroutine.

\section{Conclusions}

We have derived lower bounds on the number of time steps for
simulating arbitrary pair-interactions between $n$ qudits.  Like the
lower bounds on the time overhead, they make use of the spectrum of
the coupling matrices. However, there is no direct connection between
both complexity measures since the time overhead refers to spectral
majorization while the bounds on the number of time steps refer to the
number of eigenvalues not contained in a certain interval.  We have
shown an example where the number of time steps is of the order $n$
but the time overhead is independent of $n$.

\section*{Acknowledgements}

Thanks to Debbie Leung for helpful corrections.
This work has been founded by the BMBF project 
``Informatische Prinzipien 
und Methoden bei der Steuerung komplexer Quantensysteme''.

%
%



\end{document}